\documentclass[12pt, letterpaper]{article}

\usepackage[utf8]{inputenc}
\usepackage[margin=1in]{geometry}
\usepackage{booktabs}

\usepackage[labelfont=bf, skip=5pt, font=small]{caption}

\usepackage[]{graphicx}

\usepackage{natbib}

\usepackage{amsmath,amssymb}

\usepackage{textcomp} 

\begin{document}

\begin{center}
\huge{Self-Organized Criticality Explains \par Readiness Potential}
\par
    \large{Katsushi Kagaya}\footnote{
    corresponding to kkagaya@mail.kitami-it.ac.jp, Faculty of Engineering, 
    National University Corporation Hokkaido Higher Education and Research System, Kitami Institute of Technology,
    165 Koen-cho, Kitami, Hokkaido, Japan
    },
    \large{Tomoyuki Kubota}\footnote{
    kubota@isi.imi.i.u-tokyo.ac.jp,
    Next Generation Artificial Intelligence Research Center (AI Center),
    The University of Tokyo, 7-3-1 Hongo, Bunkyo, 113-8656, Tokyo, Japan
    },
    \large{Kohei Nakajima}\footnote{
    k-nakajima@isi.imi.i.u-tokyo.ac.jp,
    (1) Graduate School of Information Science and Technology, The University of Tokyo, 7-3-1 Hongo, Bunkyo, 113-8656, Tokyo, Japan
    (2) Next Generation Artificial Intelligence Research Center (AI Center)
    }
    
\end{center}

\begin{abstract}
  Readiness potential is a widely observed brain activity in several species including crayfish before the spontaneous behavioral initiation. However, it is poorly understood how this spontaneous activity is generated.
  The hypothesis that some specific, dedicated site is responsible for the spontaneity has been questioned. 
Here, by using intracellular recording and staining of the brain neurons in crayfish and modeling using the sandpile, which is the original model of self-organized criticality (SOC), we show that readiness potential can emerge everywhere in the brain because it is a SOC system.
Despite the diversity in neurons and their morphology, brain neurons showed signatures of criticality and readiness potential.
We find that the previously known readiness potential in a neuron is a consequence of the critical behavior of the entire network.
Indeed, seemingly unrelated membrane potential activity in neurons in different animals can shape readiness potential when its time series are averaged after their alignment with respect to the spontaneous behavioral initiation.
We show that the sandpile model not made for the potential, can form the premovement buildup activity similar to readiness potential.
Scaling properties of the synaptic avalanches are in line with those of vertebrate species; thus, not only is the critical brain hypothesis supported in crayfish, but our findings might also provide a unified view of the basis of spontaneity in animal behavior.

\end{abstract}

\pagebreak
\section*{Introduction}

Readiness potential first recorded in human brain via an electroencephalogram (\cite{kornhuber1965changes}).
However, not only in the human brain (\cite{kornhuber1965changes,schurger2012accumulator,travers2021readiness}), but also the potential were recorded in rats (\cite{murakami2014neural}), birds (\cite{daliparthi2019transitioning}), fish (\cite{ramirez2021ramp}), and crayfish (\cite{kagaya2010readiness,kagaya2011sequential}).
In crayfish, it can be recorded in a single neuron via intracellular recording (\cite{kagaya2011sequential}).
The presynaptic bombardments and anatomical features of the neuron suggest that readiness potential is generated at the network level, not at the single neuron nor single sites in the brain.
Thus, the ubiquity of the readiness potential ``from crayfish to human''(\cite{schurger2012accumulator}) suggests some general principle, as physicists have previously pointed in a different context when proposing the concept of self-organized criticality (SOC) in lobster and human brains (\cite{bak1987self, chialvo1999learning}).
Our goal in this paper is to explain the readiness potential through the concept of SOC with neurophysiological evidence.
The most unclear point of readiness potential is how ``spontaneous'' activity is generated.
One hypothesis is that there is one specific site that is obligated to trigger initial process of the premovement potential.
This hypothesis will lead to the next question: Which site is the presynaptic trigger of the specific site, and so on?
Here, we newly provide an evidence from an alternative view; every site in the critical state can trigger premovement activity.
A similar interpretation has previously been provided based on a leaky accumulator model including a stochastic term (\cite{schurger2012accumulator,schurger2015nowhere}).
In contrast, we provide the view of dynamical systems: SOC.

\begin{figure}[t!]
        \centering
                \includegraphics[width=\textwidth]{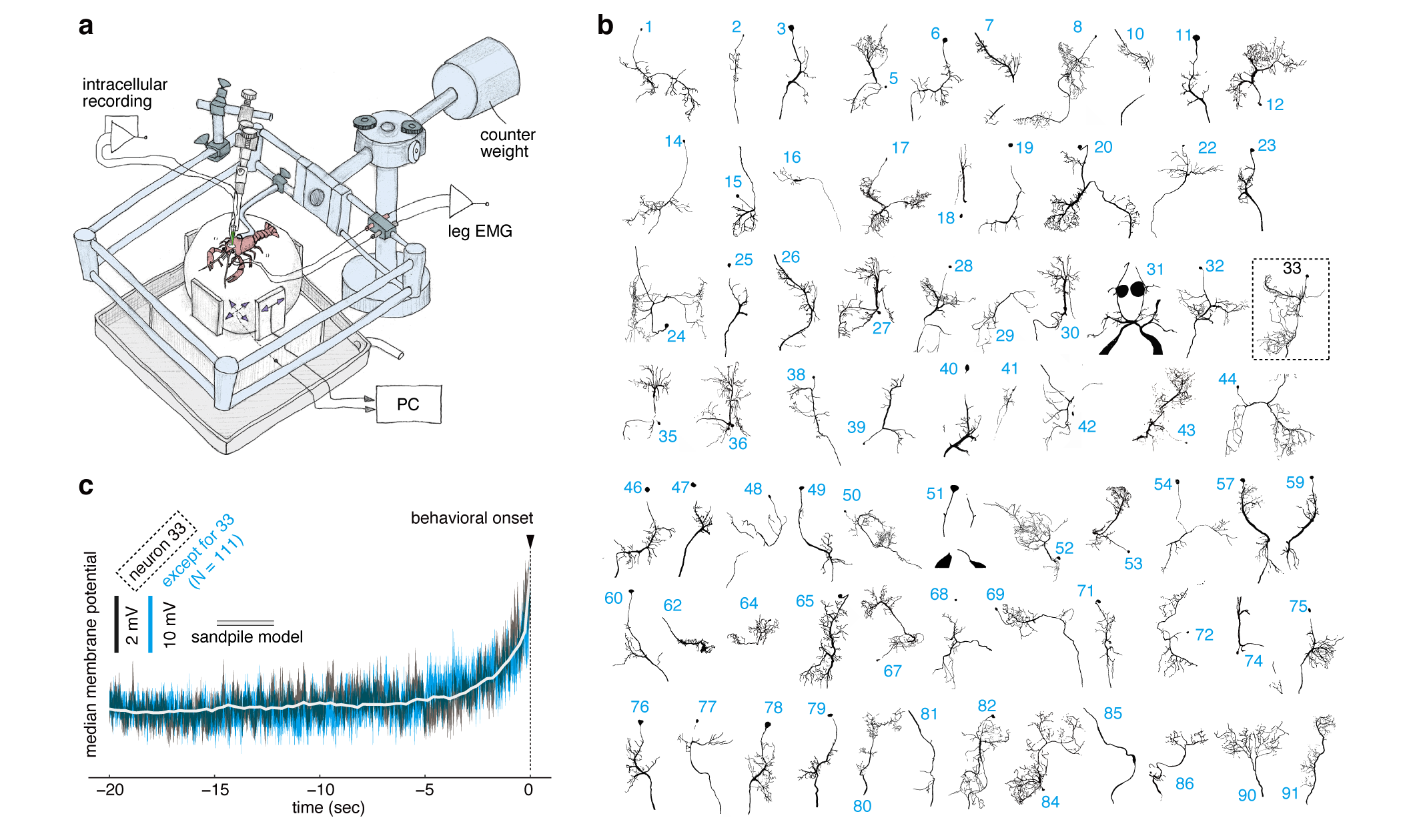}
                \caption{ 
                Description of readiness potential from a single neuron by other virtual neuron population and sandpile model.
                \textbf{a,} The membrane potential activity was examined by intracellular recording. Tethered crayfish can start walking on the spherical treadmill of their own accord. The initiation timing was determined by leg electromyography. 
                \textbf{b,} Morphology of brain neurons were successfully stained and recorded sufficiently long time.
                Neuron \#33 showed readiness potential (\cite{kagaya2011sequential}). 
                \textbf{c,} The overdrawn readiness potential of neuron \#33 compared to that of the virtual population of other neurons from different individual crayfish, and the sandpile model.
                The behavioral onset is defined by the activation of the leg electromyography in (\textbf{a}). 
                The median value of the recorded membrane potential (\textbf{a}) is presented. 
                The sandpile activity (supplementary material) is scaled in the both x and y axes.
                }
                \label{fig:method}
\end{figure}

The theory of quasi-criticality has been proposed as a realistic explanation of critical neural activity in vertebrates (\cite{williams2014quasicritical,fosque2021evidence,mariani2022disentangling,obyrne2022critical}).
It can provide us with the framework to examine the crayfish brain.
Studies of vertebrates have been performed by recording of extracellular spikes and local field potentials. 
In this study, we investigate whether the crayfish brain shows critical behavior through the intracellular single-cell study of the neurons in the brain.
To this end, three signatures are required to demonstrate whether a system is critical (\cite{helmrich2020signatures,fosque2021evidence}); 
(1) multiple power-laws in duration, size, and spectrum of the activity; 
(2) exponents of the duration and size of avalanches satisfying the crackling noise relation (\cite{sethna2001crackling,jensen1998self}); 
(3) a universal scaling function characterizing the scale-invariant shape of avalanches (\cite{friedman2012universal}).
The crackling noise relation (\cite{sethna2001crackling,jensen1998self}) is satisfied when the two exponents of size and duration of avalanches, $\alpha$ and $\tau$, satisfy $(\tau-1)/(\alpha-1)=\gamma$, where $\gamma$ is the slope of the log-log plot of the size and duration of the synaptic activity.
Furthermore, we compare the scaling relation of the neurons with the BTW sand pile (\cite{bak1987self,bak1988self}), the stochastic model (leaky accumulator) proposed for readiness potential (\cite{schurger2012accumulator}), and the real data extracellularly recorded in vertebrate species (\cite{fontenele2019criticality,fosque2021evidence}).

\section*{Materials and methods}
\subsection*{Animals and electrophysiology}
This study overlaps with a previous study (\cite{kagaya2011sequential}), but contains new data and updated analyses. 
A more detailed morphological description of brain neurons, including newly identified ones is provided. 
The crayfish \textit{Procambarus clarkii} of both sexes with body length of 8--12 cm were sacrificed for the experiments. 
We additionally included long spontaneous neural activity when the animal did not walk. 
The number of individual was $N=161$ in total. 
A total of 47 crayfish were excluded based on the quality of the recording, but all data including them are also available in the supplementary dataset (https://doi.org/10.6084/m9.figshare.26132065).
To determine when the animal was walking, electromyography from the mero-carpopodite flexor in the second walking leg was obtained using a pair of wire electrodes (Teflon-coated silver wires with a diameter of 125 {\textmu}m). 
One recording from one leg was sufficient to determine when the animal started walking with an error up to 0.5 s, because all eight legs were found to be activated within the time range of 65.89 $\pm$ 6.40 ms (\cite{chikamoto2008electromyographic}). 
The animals were tethered to the treadmill (Fig. \ref{fig:method}a) with electrodes and holding manipulators, but they could stand up and down with their own weight. This was achieved by a counterbalanced weight (\cite{kagaya2010readiness}). 

The glass micro-electrodes, filled with 3{\%} solution of Lucifer yellow in 1 M lithium chloride (30--50 $M{\Omega}$), were penetrated to the dendrites of the single brain neurons. 
Brains contaminated with more than two neurons were excluded.
Thus every membrane potential activity obtained from one neuron was from one individual crayfish.
After recording was completed, the cells were stained with Lucifer yellow by injecting hyperpolarizing current pulses. 
The brain was dissected, fixed in formalin, dehydrated in an alcohol series, and cleared in methyl salicylate. 
The neurons were optically sectioned using a confocal laser scanning microscope. 
The 2D dendritic morphology of each neuron was manually traced and digitized. The 3D morphology was reconstructed in ImageJ (https://imagej.net/ij/). 

\subsection*{Power spectral analysis}
To obtain long-range time series for the power spectral analyses, we selected the animals for which the stable intracellular recordings were successful ($N=92$, Fig. \ref{fig:power_spectra}a). 
We separately extracted the time series when the animal was in a silent phase for the analysis of the shape of the synaptic activations.
The synaptic avalanches were defined by size and duration, power-law fitting to a histogram was performed.  
Each setting can be confirmed in the supplementary codes and plots. 
The number of bins of histogram was set 50 or 100.
The region of fitting was determined visually for each histogram. 
No quantitative criterion was used to determine whether the crackling noise relation (\cite{sethna2001crackling,jensen1998self}) was satisfied. 
However, the validation can be confirmed in the plots of data and predicted lines.
The complete information was provided in the supplementary data. 
The codes for the analyses can also be found in the directory. 

\subsection*{Making of readiness potential}
The readiness potential is not an actual, single time series of neural activity.
In general, it is constructed by time-reversely averaging several time series aligned with the time points of voluntary initiation of movement (\cite{schurger2021readiness}).
The neural readiness potential was constructed by aligning the membrane potential $V_{m}$ with the behavioral onsets determined from the leg electromyography.

Similarly, we ran the sandpile model with cells with 100 $\times$ 100, following the canonical setting of the BTW sandpile (\cite{bak1987self,jensen1998self}). 
The degree of ``the activation of the sandpile'' was defined by the total number of the overcritical sites (more than 4) for a single time step. 
The initial subcritical series were removed, and only the data after the system became critical were used to record the time series of the activation. 
The threshold to the onset was set to 99 percentile of the largest avalanche. 
Even when we changed the threshold value to 85, 90, and 95 percentiles, the shapes of the readiness potentials did not change when they were appropriately scaled (SFig. \ref{fig:sandpile_readiness}).

\subsection*{Scaling function}
The scaling function representing the shape of avalanches, was examined by scaling the duration and the size of avalanches. 
The scaling functions for the neurons and models (the sandpile and leaky accumulator model) were examined.
To focus on the synaptic activity, we excluded the time series that showed strong spike activity.
The neural time series were high-pass filtered more than 1 Hz to remove the slow trend.
The leaky accumulator model that we used here was proposed for describing the readiness potential (\cite{schurger2012accumulator}).
The equation is the same with the paper: 
\begin{equation*}
        \delta x_i = (I-kx_i) \Delta t + c \xi_i \sqrt{\Delta t},
\end{equation*}
where $I=0.11$, $x_0=0$, $k=0.5$, $\delta t=0.001$, and $c=0.1$, $\xi_i \sim Gaussian(0,1)$. 
We used this setting to generate the time series, and we performed an analysis of the scaling relation by extracting the scaling function from another time series of synaptic and sandpile activity.
The avalanches were classified into four classes using 35, 50, and 80 percentiles of durations.
The mean values along the scaled time axes for each class were calculated to generate a single wave as a shape of the avalanche.
The four waves were overlaid to ensure that they matched each other. 
If they completely matched, the single wave form was called a universal scaling function. 


\section*{Results}
\subsection*{Readiness potential in a neuron, neurons, and sandpile}

\begin{figure}[!t]
  \centering
      \includegraphics[width=\textwidth]{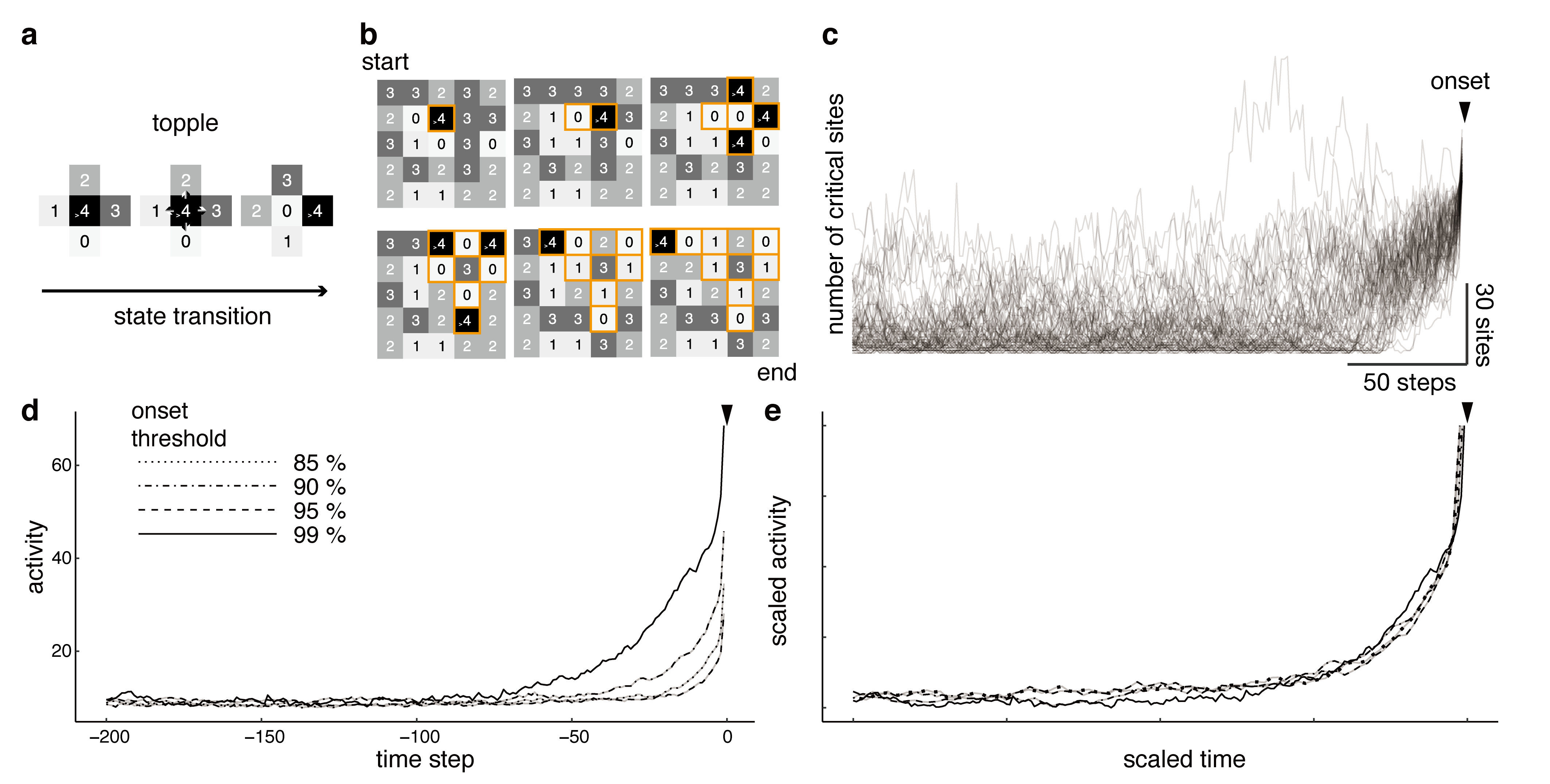}
      \caption{
        \textbf{(supporting material)} Unchanged shape of sandpile readiness potential when the decision threshold is changed.
        \textbf{a,} Rule of distribution of sand grains. Each cell contains sand grains and the number of grains is represented by the numbers in the cell.
        When the number of grains becomes larger than four, the four grains are distribute to the four neighbours.
        \textbf{b,} One example of an avalanche. The avalanche starts from a site that becomes more than four. This avalanche ends at a size of 10 ($10 = 1+1+3+3+1+1$) and duration of 6.
        \textbf{c,} We consider the number of critical sites for each time step as ``activation'' of the sandpile. The activation series are aligned with the timing of when the activation becomes 99 percentile of the avalanches. The series are overdrawn.
        \textbf{d,} The mean value of the series are shown when the four different thresholds are used.
        \textbf{e,} The four different activities can match each other when the time and activity are scaled.
      }
  \label{fig:sandpile_readiness}
\end{figure}

The crayfish were tethered to the frame but allowed to walk freely on the treadmill (Fig. \ref{fig:method}\textbf{a}). 
No sensory stimuli were provided to trigger walking.
We determined when such spontaneously initiated walking occurred from the electromyographic (EMG) activity.
Simultaneously, we intracellularly recorded the membrane potential and stained neurons to reveal their morphology using the sharp glass micro-electrodes (Fig. \ref{fig:method}\textbf{b}).
The walking bouts were found to be preceded by the following two time series: 
(1) the averaged membrane potential of the neuron \#33;
(2) the averaged membrane potential of all the neurons except the neuron \#33.
These two time series showed the same shape when appropriately scaled (Fig. \ref{fig:method}\textbf{c}). 
Furthermore, we confirmed that (3) the averaged ``activity'' of the sandpile could also form the same shape of premovement buildup activity.
For (1) and (2), we scaled only the membrane potential while the time was unchanged. 
For the sandpile activity (3), we scaled both axes.
Even when we changed the threshold values, the shape of buildup activity was unchanged when appropriately scaled (SFig. \ref{fig:sandpile_readiness}).
Thus, in addition to the previously reported fact (1) (\cite{kagaya2011sequential}), the shape of the readiness potential was well described by the other neurons (2) as well as the original sandpile model (3) proposed to instantiate the concept of SOC. 

\subsection*{Power spectra of the brain neurons}
\begin{figure}[!t]
        \centering
            \includegraphics[width=0.75\textwidth]{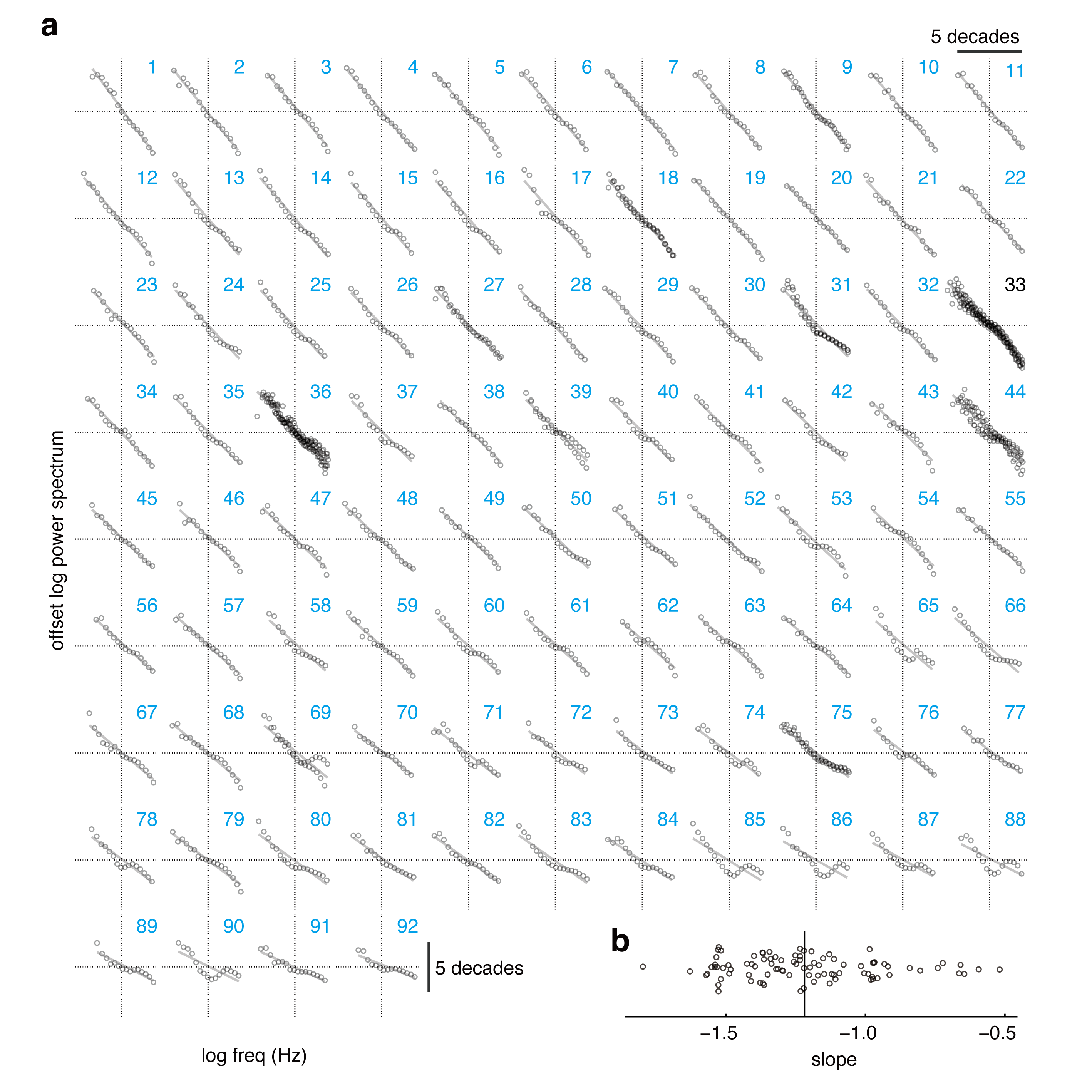}
                    \caption{ 
                Common power spectra of the diverse brain neurons ($N=92$).
                    \textbf{a,} Despite the morphological diversity shown in the previous figure, the neurons show a common power-law in the power spectra. The power law fitting is up to six decades in several cases. 
                    The fitting region is not limited to the linear part, but includes the bumps at the higher frequency region. The bumps are likely due to the spike activity. 
                    \textbf{b,} Slopes of the neurons distributed around -1.22 are indicated by a vertical line. 
                    }
                    \label{fig:power_spectra}
\end{figure}

\begin{figure}[!t]
\centering
        \includegraphics[width=\textwidth]{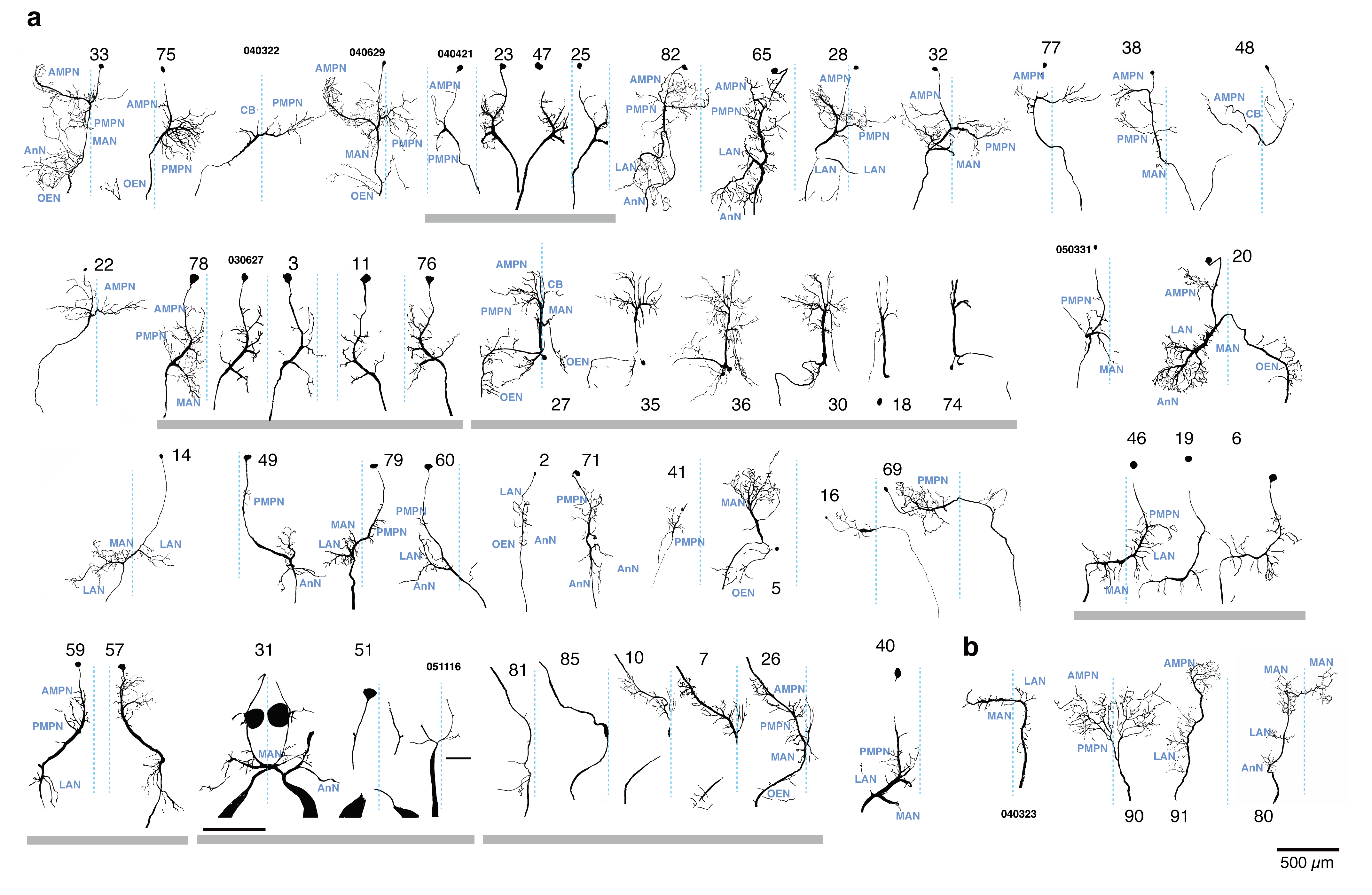}
        \caption{
                \textbf{(supporting material)}
                Descending and ascending neurons successfully stained.
                \textbf{a,} Descending neurons that extend the axon to the subesophageal and thoracic ganglia. Their cell bodies resides in the brain. The neurons \#81, \#85, \#10, \#7, and \#26 are listed as putative descending neurons because no cell bodies were observed. 
                The neuron \#33, \#75 \#27 are described as ``readiness discharge'', ``continuation'', and ``termination'' neurons recruited for voluntary walking (\cite{kagaya2011sequential}). 
                The neuron \#040421, \#23, \#47, and \#25 are described in a previous work related to sensory information processing of the statocyst (\cite{hama2003effects}). 
                Neuron \#20 is described as ``S5'' (\cite{nakagawa1989morphology}). 
                Neurons \#46, \#19, \#6 are described as ``\#8'' (\cite{hama2003effects}).
                Neuron \#59, \#57 are described as ``\#S3'' (\cite{nakagawa1989morphology}) and as ``\#4'' (\cite{hama2003effects}).
                \textbf{b,} Ascending neurons that project axon collaterals in the brain.
                The neurons with the underlines have similar dendritic structures, and are used in Fig. \ref{fig:functional_diff}. The numbers are assigned based on the spectral analysis in Fig. \ref{fig:power_spectra}. The names of brain regions are as follows. AMPN: anterior medial protocerebral neuropil. PMPN: posterior medial protocerebral neuropil. MAN: median antenna I neuropil. LAN: lateral antenna I neuropil. OEN: esophageal neuropil. 
                The nomenclature is based on the report of other decapod crustaceans \cite{sandeman1992morphology}.
                The broken blue lines indicate the lines dividing the left and right sides of the central nervous system.
        }
        \label{fig:des_asc}
\end{figure}

\begin{figure}[!t]
\centering
        \includegraphics[width=\textwidth]{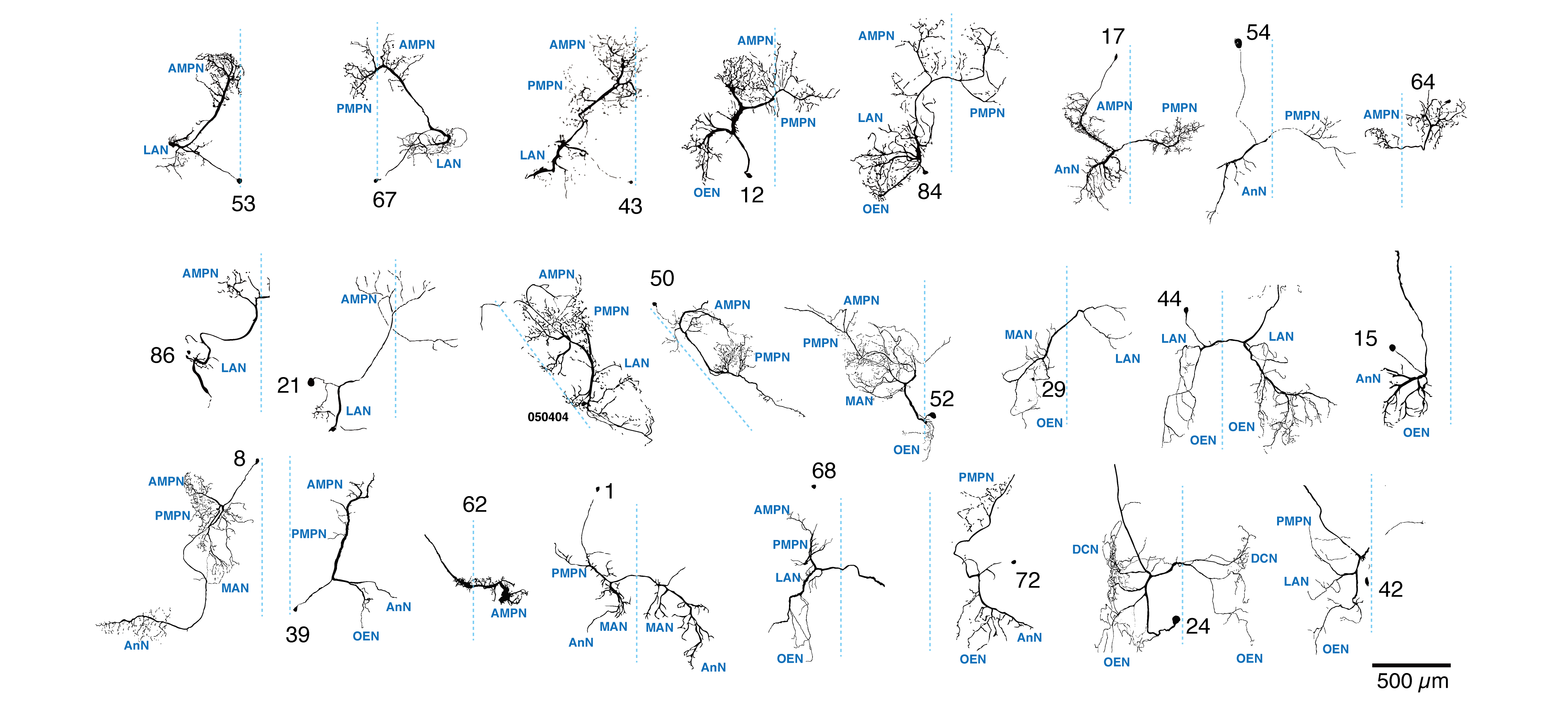}
        \caption{
                \textbf{(supporting material)}
                Local neurons successfully stained. Cell \#62 is a non-spiking giant interneuron (NGI). 
                The names of the brain regions are as follows.
                AMPN: anterior medial protocerebral neuropil. PMPN: posterior medial protocerebral neuropil. MAN: median antenna I neuropil. LAN: lateral antenna I neuropil. OEN: esophageal neuropil. 
                The nomenclature is based on the paper (\cite{sandeman1992morphology}).
                The broken blue lines indicate the lines dividing the left and right sides of the central nervous system.
                Neuron \#12 was described as a non-spiking neuron whose synaptic activity change prior to the spontaneously initiated walking (\cite{kagaya2011sequential}). 
                Neuron \#84 was also described in the same way, but it is a spiking neuron.
                Neuron \#62 is one of NGI (\cite{okada1988nonspiking,fujisawa2007physiological}).
                Fujisawa and Takahata (2007a) define neuron \#53 as L9 and L10 and neuron \#17 as L7, and they describe these as neurons upstream to NGI.
                Neuron \#1 is defined by Hama and Takahata (2005) as a Type III neuron affecting descending neurons.
        }
        \label{fig:local}
\end{figure}

We further examined the signatures of SOC in individual neurons in the brain.
First was the 1/f fluctuation, which has been described as a ``finger print'' of SOC (\cite{bak1987self}).
The dendritic structures of the neurons encountered were summarized and labeled with the numbers (Fig. \ref{fig:method}\textbf{b}) in the order of the slopes of power-law fitting to the power spectra (Fig. \ref{fig:power_spectra}\textbf{a}). 
The areas of dendritic projections in the brain with the actual scale are described in SFig. \ref{fig:des_asc} and SFig. \ref{fig:local}. 
The number of neurons in Fig. \ref{fig:method} ($N=72$) is the subset of the number in Fig. \ref{fig:power_spectra}\textbf{a} ($N=92$), because of the lack of morphological information. 
We analyzed the total length of the time series, 11 h 41 min 26 s,  to show the spectra. 
Remarkably, the spectra in the log-log plots show linear relation up to six decades in addition to the neurons mentioned above.
Overall, the slope values of the lines were distributed around the mean -1.22 (Fig. \ref{fig:power_spectra}\textbf{b}). 
The tail parts of the spectra were much more common among the neurons; thus, the evaluation was still underscored. 
The intracellular time series in this study showed the wide power laws up to six decades.
This far surpasses the range of power law (four decades) in the time series of calcium imaging data in mice (\cite{jones2023scale}).

\subsection*{Readiness potential is not a single, special activity}
\begin{figure}[p]
        \centering
                \includegraphics[width=\textwidth]{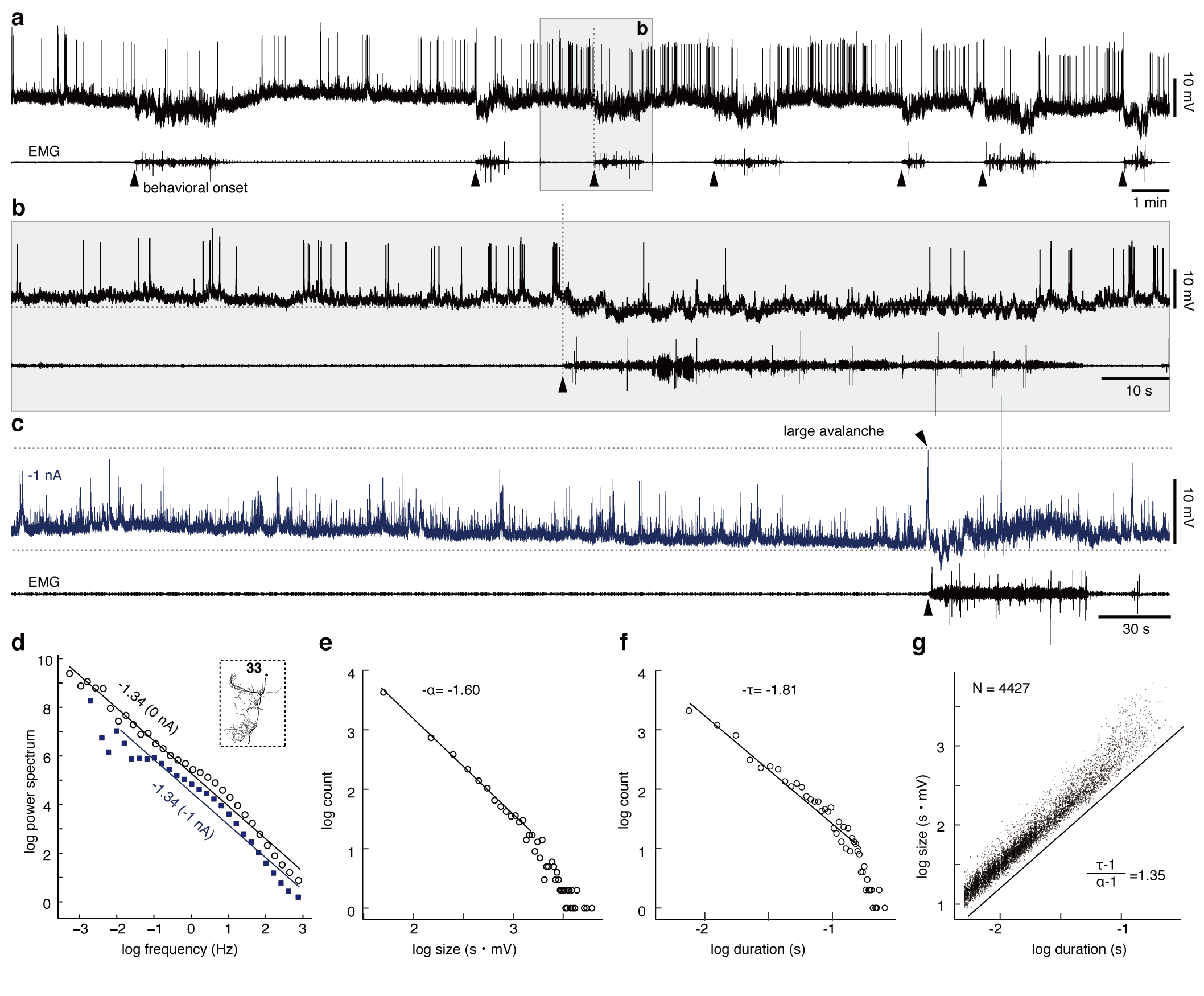}
                \caption{ Membrane potential activity of the descending neuron generates readiness potential with the synaptic activity following power law.
                \textbf{a,} Upper trace: continuous, long, intracellular synaptic activity with the electromyography. Lower trace: electromyography from a walking leg. Seven walking bouts can be observed.
                \textbf{b,} The expanded part around an event of the spontaneous initiation of walking indicated by the rectangle in \textbf{a}.
                \textbf{c,} Hyperpolarized current was injected so that the spikes were suppressed to experimentally emphasize the synaptic activation. 
                A large synaptic summation was observed just before the behavioral onset. 
                \textbf{d,} The power spectrum of the neuron. 
                Open circles are the results from the upper trace in \textbf{a}. 
                The closed dark blue circles are from the upper trace in \textbf{c}.
                The power law is robust up to about six decades. 
                The current injection did not break the power law. 
                \textbf{e,} The power-law distribution of the sizes (overthreshold area of the activation after removing the slow decreasing trend) of the synaptic activations extracted from \textbf{c}. 
                \textbf{f,} The power-law distribution of the durations of the activations.
                \textbf{g,} The joint distribution of the sizes and durations of the avalanches.
                The crackling noise relation using the values $\alpha$ and $\tau$ is well satisfied; the slope of dots and the line are well matched.
                }
                \label{fig:readiness_potential}
\end{figure}

Readiness potential is usually presented as one single averaged time series that is aligned with the timing of initiation.
This is because we are interested in the averaged time series just before the spontaneously initiated movements.  
In contrast, here we presented the example of long time series with several walking bouts (Fig. \ref{fig:readiness_potential}\textbf{abc}) to examine to what degree of power law holds.
To eliminate the spike activity of the time series, the hypterpolarizing current was injected (Fig. \ref{fig:readiness_potential}\textbf{c}).
Even when we remove the spike activity, the power spectra of the time series were unchanged (Fig. \ref{fig:readiness_potential}\textbf{d}).
It should be noted that the large synaptic activation was observed just prior to the spontaneously initiated walking (Fig. \ref{fig:readiness_potential}\textbf{c}).

We examined the distribution of the sizes and durations of synaptic activations (Fig. \ref{fig:readiness_potential}\textbf{efg}).
The distributions in log scales demonstrate that the large activation is just one example of the common power-law distribution.
Furthermore, the relation of exponents was well satisfied; the slope of the dots was almost the same with the line calculated from the exponents (Fig. \ref{fig:readiness_potential}\textbf{g}).
Thus, the activations from small to large sizes and durations were well described by the single straight line.

\subsection*{Similar scaling of crayfish and vertebrates}
\begin{figure}[!t]
\centering
        \includegraphics[width=0.5\textwidth]{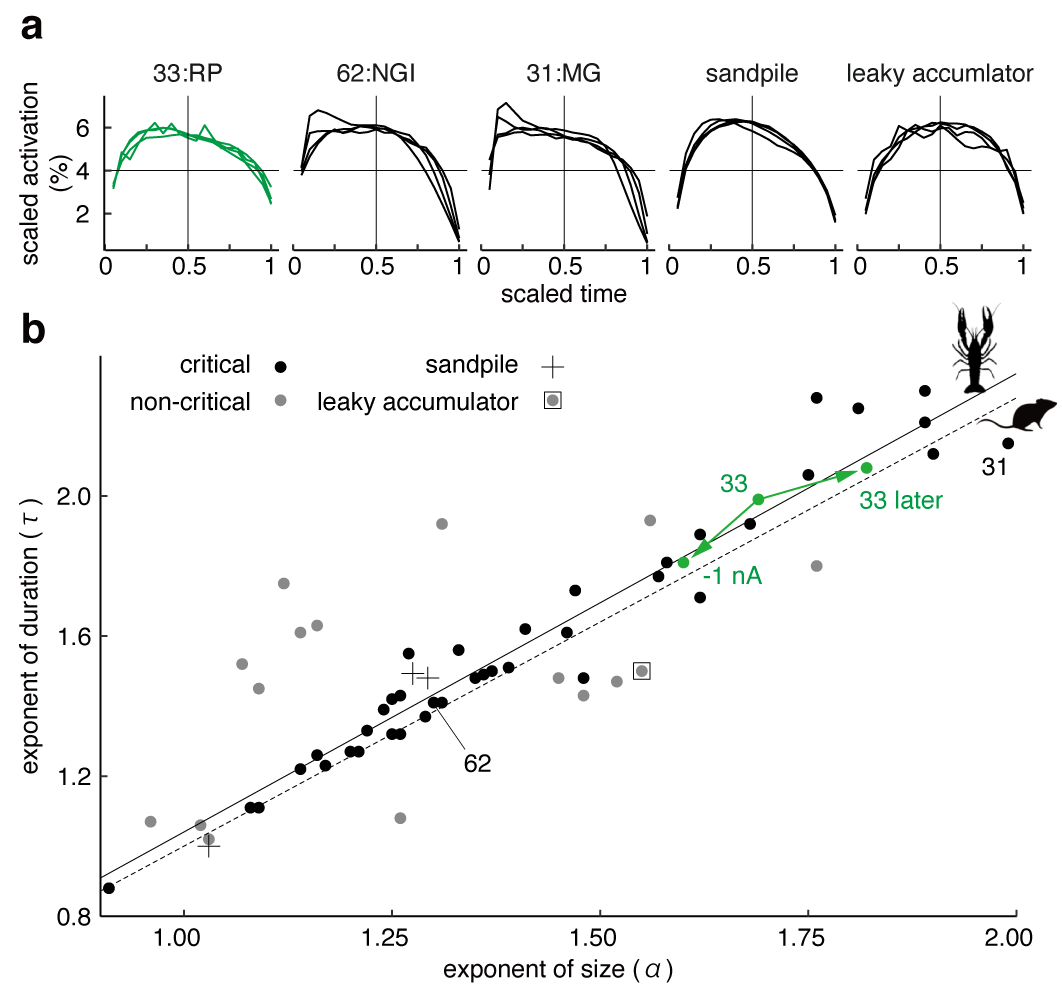}
        \caption{
        Universal scaling functions and scaling relation of neural, model avalanches.
        \textbf{a,} 
        Examples of universal scaling functions. 
        Four shapes from different sized domains of durations of avalanches match.
        The deviations of neurons \#62 (NGI) and \#31 (MG) around 0 may be due to finite sampling effect.
        \textbf{b,} 
        The linear relation of the exponents $\alpha$ and $\tau$. 
        The slope of the broken line is from rat data; slope is $1.28$ (\cite{fontenele2019criticality}). 
        The slope of the fitted line to the black dots indicating critical neurons in crayfish is $1.31$. 
        Neuron \#33 that showed readiness potential (RP, shown in green dots) hovers around the line even under current injection or after about 30 min. 
        Ten neurons including three MGs were omitted.  
        }
        \label{fig:scaling}
\end{figure}

The scaling relation of the synaptic activation in crayfish was analyzed (Fig. \ref{fig:scaling}\textbf{ab}) to compare the results with other animal species.
Also, we compared the results with the models including the sandpile (\cite{bak1987self,jensen1998self}), stochastic leaky accumulator models (\cite{schurger2012accumulator}).
Neurons with too many spikes were removed for the analysis.
Then, 76 neurons were examined. 
The synaptic activity when the animals were silent was analyzed to focus on the stable activity. 

Avalanches of four classes using 35, 50, and 80 percentiles of duration were overdrawn after scaling, as seen in Fig. \ref{fig:scaling}\textbf{a}.
In either case, we successfully extracted universal scaling functions that characterize the shape of the avalanches (Fig. \ref{fig:scaling}\textbf{a}).
Except for the deviations in the two smaller classes of the non-giant interneurons NGI (\#62, \cite{okada1988nonspiking,fujisawa2007physiological}) and the medial giant neurons MG (\#31, \cite{wiersma1947giant}), the successful matches of the shapes were confirmed.
Only the stochastic leaky accumulator model showed a symmetric shape; the other neurons and sandpile model showed asymmetric shapes.

Furthermore, we mapped the exponents of the sizes and durations to compare the avalanche shapes with those of other species (Fig. \ref{fig:scaling}\textbf{b}).
The slope of the relation of the critical neurons that satisfied the scaling relation was found at $1.31$ for crayfish and close to $1.28$ for rats, mice, turtles, monkeys (\cite{fontenele2019criticality}). 
The leaky accumulator model did not satisfy the criterion.

Although the position of the points of \#33 (green dots in Fig. \ref{fig:scaling}\textbf{b}) might change because of the non-stationarity, it still hovered around the line.
In addition, it stayed around the line after more than 30 min or when a hyperpolarizing current ($-1$ nA) was injected.

\section*{Discussion}
\begin{figure}[!t]
        \centering
                \includegraphics[width=\textwidth]{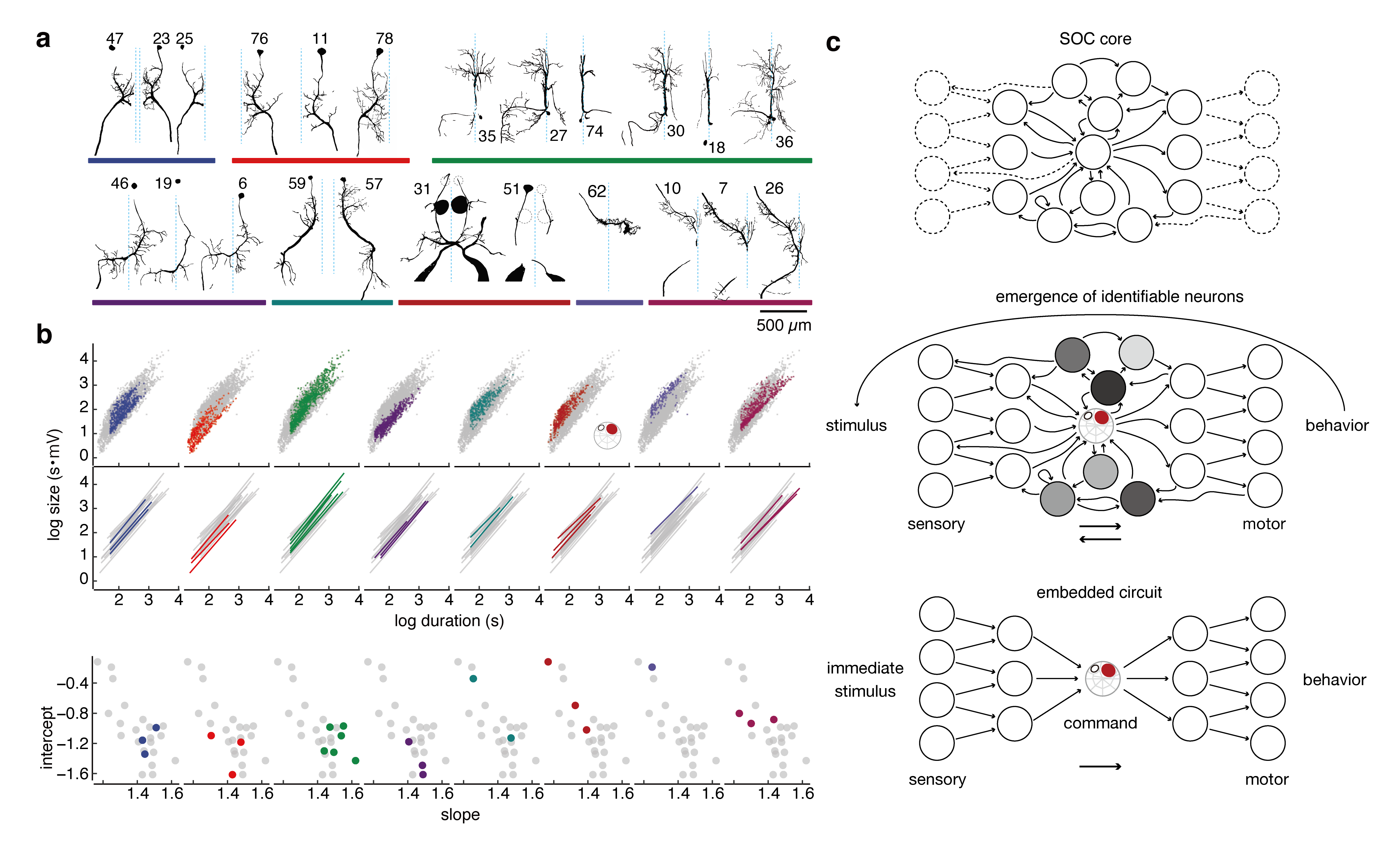}
                \caption{
                  \textbf{(supporting material)}
                Hierarchical variability in morphology and physiology of identified neurons. 
                \textbf{a,} 
                Identified neurons share common morphology yet with different micro-morphology in their neurites across individuals.
                The groups are designated with eight different colors.
                Except for the group of \#10, \#7, and \#26, all of the neurons have been reported in the literature (\cite{hama2003effects,fujisawa2007disynaptic,kagaya2011sequential,okada1988nonspiking}).
                \textbf{b,}
                The relation of the duration and size of the synaptic avalanches in the log scales.
                Upper dots with colors were randomly sampled ($N=200$) for each neuron. 
                The lines in the middle row with colors are fitted to the dots for each neuron. 
                The bottom plot shows the relation of the slope and intercept of the lines.
                \textbf{c,}
                The schematic diagram summarizing our picture of this study. 
                We can recognize that the hierarchical characterizations in both morphology and dynamics, but the neurons are in critical states.
                We speculate that the hierarchical variability would be a fingerprint of SOC at extended scales in the evolutionary and developmental time scales in addition to the physiological scales associated with the readiness potential reported in this study.
                }
                \label{fig:functional_diff}
\end{figure}

Behavioral initiation has long been discussed in the dichotomy between spontaneous or reactive (e.g., \cite{tinbergen1951study}), but it is unified in the view of this study. 
So far, ``readiness potential'' is a term specific to the potential activated prior to the spontaneous initiation of a behavioral act.
Our results demonstrate that the spontaneous activation of a large readiness potential is an example of the same power-law distribution as those of small synaptic activation without an apparent association with behavior.
Meanwhile, the reactive escape responses of crayfish is mediated by the command neurons and other non-giant neurons (\cite{wine1972organization,edwards1999fifty}). 
The reactive or functional neurons have often been examined as ``identified neurons'' (see the examples encountered in this study, SFig. \ref{fig:functional_diff}\textbf{a}).
We found that their synaptic activities also take place in the critical states.
We further found that their synaptic activity has slight modifications for each identified neuron (SFig. \ref{fig:functional_diff}\textbf{b}); similar neurons are likely to share similar scaling relations.
Although this point requires further physiological study for each identified neuron, overall, the brain neurons are in critical states.
The question is why the reactive neurons are also in critical states.
It is hypothesized that although these neurons require sensory inputs to be functional, the inputs will be summed up on the ongoing synaptic activity maintained in the critical states.
In other words, the variability of sensory motor responses would be due to the critical synaptic activity so far referred as ``background ongoing activity''.
Now that the activity is not background anymore, we consider that the spontaneous synaptic activity is exploited for either spontaneous readiness potential or adding variation to the sensory responses.
Thus, the two extremes---spontaneous and reactive behaviors---are now unified in the same picture.
They are both constrained to be ``ready'' in criticality.
Further detailed study on the integration of critical activity and multiple sensory information is needed. 

Although our study is limited to ``spontaneous activity'',  the scaling relation has been found to be close to vertebrate species, suggesting that the generality of this view can extend to other species.
Our findings on the scaling relation may further offer insights into other invertebrate and vertebrate systems. 
No such reports on the membrane potential activity have been seen so far, but spontaneous behavior at the individual level has been well described as power-law distributions in many species (\cite{martin1999temporal,maye2007order,reynolds2018current, abe2020functional,arata2022insulin}). 
For example,  power-law behaviors, including the L\'{e}vy walks, have been well known in \textit{Drosophila melanogaster} (\cite{sims2019optimal,abe2020functional}) and discussed in the context of behavioral ecology (\cite{bartumeus2016foraging}) and voluntary behavior (\cite{maye2007order}). 
The observation of neuronal avalanche in the invertebrate nervous system is still rare, except for the extracellular study in leeches (\cite{garcia2007spontaneous}). 
It is expected to be confirmed whether the intracellular neural activity might show the neuronal avalanche that can be observed as the readiness potential.
Our study implies that the readiness potential in vertebrates may also have this common basis. 

In conclusion, we have shown that the readiness potential is well described by the network dymics at the critical point.
The critical dynamics are not likely to be generated in some specific neurons or site, but all of the neurons are more or less critical and involved in shaping readiness potential.
The crayfish nervous system has many functional neurons with diverse morphology, but criticality is the core of spontaneity, and specific functionality is built on this core.
Readiness potential can emerge everywhere in the network, like a sandpile can generate large avalanches everywhere.

\bibliographystyle{plainnat}
\bibliography{ref}

\section*{Acknowledgments}
We thank Dr. Masakazu Takahata for the electronic work, supervising the experimental procedure, and feedback comments on the manuscript. We thank the RC seminar group (https://www.kohei-nakajima.com/rc-seminar-group) for their fruitful feedback comments.

\section*{Funding}
This work was supported by the project JPNP16007, commissioned by the New Energy and Industrial Technology Development Organization (NEDO); and by JSPS KAKENHI grants, 
JP18K19336, 
JP18H05465, 
JP19H05330, 
JP20370028, 
JP15657018, 
JP10187101. 

\section*{Declarations}

We declare no competing interests.

\newpage

\newpage

\begin{figure}[b!]
        \centering
                \includegraphics[width=\textwidth]{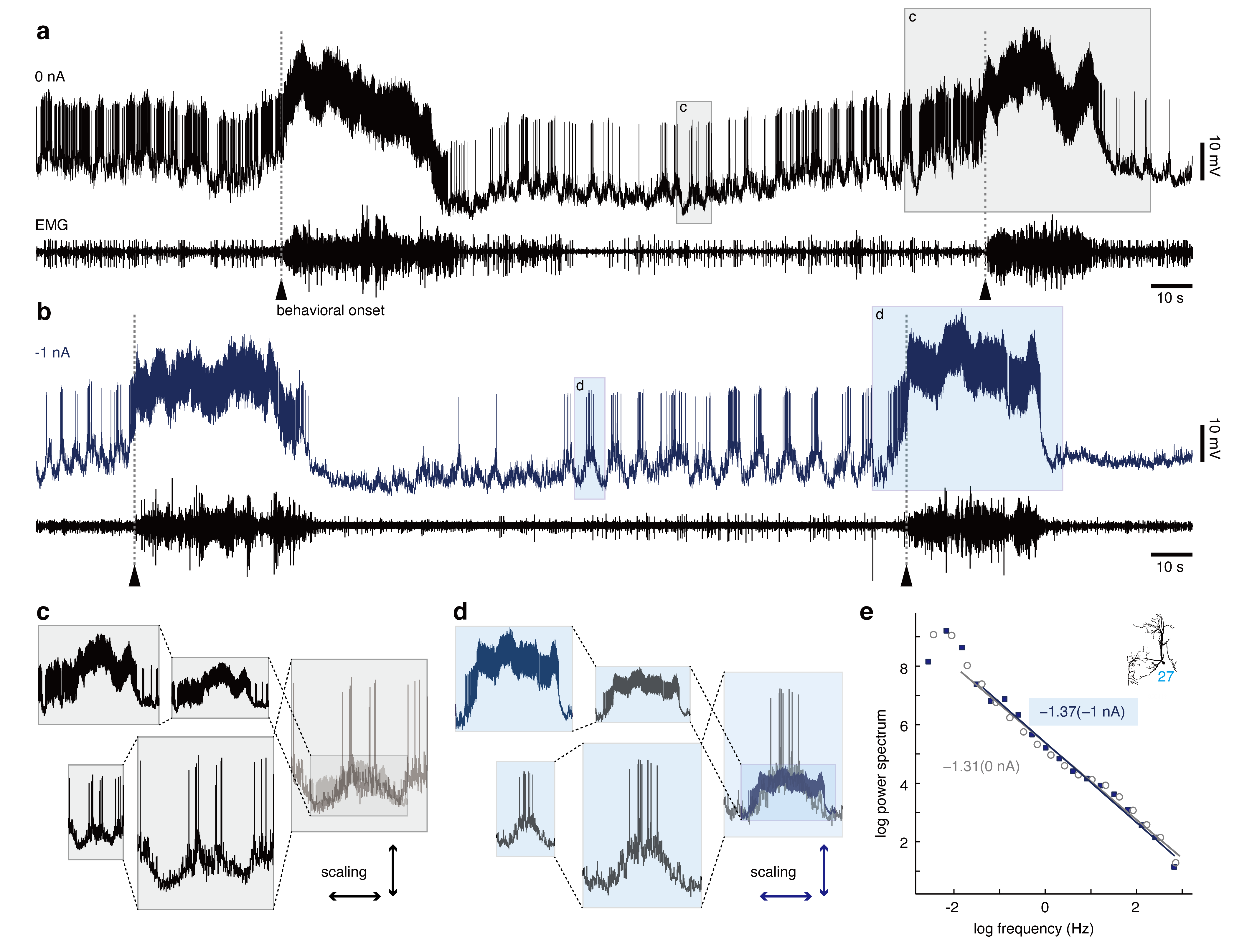}
                \caption{
                  \textbf{(supporting material)}
                \textbf{a,} Self-similarity in synaptic activity in the  descending neuron \#27. 
                The boxes are shown in \textbf{c} after changing scales along time and voltage.
                \textbf{b,} The dynamics when a hyperpolarized current ($-1$ nA) are injected so that the spike activity is suppressed.  
                The boxes are shown in \textbf{d} with scale modifications.
                \textbf{c-d,} Changing scales indicated by vertical and horizontal arrows that can fit the two time series from the traces framed by rectangles in \textbf{a} and \textbf{b} qualitatively demonstrate self-similarity. From left to right, the rectangles are changed manually. To the right, the two rectangles are overlaid to show that the shapes match. 
                \textbf{e,} The power law in power spectrum ranging up to almost five decades is robust to the current injection. Open circles are the results from the upper trace in (\textbf{a}); closed rectangle dots are from (\textbf{b,}) This shows robustness against the current injection.
                }
                \label{fig:self-similarity}
\end{figure}

\newpage

\newpage

\newpage

\end{document}